\documentclass[acmsmall]{acmart}

\AtBeginDocument{%
  \providecommand\BibTeX{{%
    \normalfont B\kern-0.5em{\scshape i\kern-0.25em b}\kern-0.8em\TeX}}}

\usepackage{soul}
\usepackage{tfrupee}
\usepackage[normalem]{ulem}

\newtheorem{hyp}{Hypothesis}

\begin{document}

\setcopyright{acmcopyright}
\acmJournal{PACMHCI}
\acmYear{2020} 
\acmVolume{4} 
\acmNumber{CSCW2} 
\acmArticle{163} 
\acmMonth{10} 
\acmPrice{15.00}
\acmDOI{10.1145/3415234}

\title{Conceptual Metaphors Impact Perceptions of Human-AI Collaboration}


\author{Pranav Khadpe}
\affiliation{%
  \institution{Stanford University}
  \city{Stanford}
  \state{California}
  \country{USA}}
\email{prkhadpe@cs.stanford.edu}

\author{Ranjay Krishna}
\affiliation{%
  \institution{Stanford University}
  \city{Stanford}
  \state{California}
  \country{USA}}
\email{ranjaykrishna@cs.stanford.edu}

\author{Li Fei-Fei}
\affiliation{%
  \institution{Stanford University}
  \city{Stanford}
  \state{California}
  \country{USA}}
\email{feifeili@cs.stanford.edu}

\author{Jeffrey T. Hancock}
\affiliation{%
  \institution{Stanford University}
  \city{Stanford}
  \state{California}
  \country{USA}}
\email{hancockj@stanford.edu}

\author{Michael S. Bernstein}
\affiliation{%
  \institution{Stanford University}
  \city{Stanford}
  \state{California}
  \country{USA}}
\email{msb@cs.stanford.edu}

\renewcommand{\shortauthors}{Pranav Khadpe et al.}

\begin{abstract}

With the emergence of conversational artificial intelligence (AI) agents, it is important to understand the mechanisms that influence users' experiences of these agents.
In this paper, we study one of the most common tools in the designer's toolkit: conceptual metaphors.
Metaphors can present an agent as akin to a wry teenager, a toddler, or an experienced butler. 
How might a choice of metaphor influence our experience of the AI agent? 
Sampling a set of metaphors along the dimensions of warmth and competence---defined by psychological theories as the primary axes of variation for human social perception---we perform a study $(N=260)$ where we manipulate the metaphor, but not the behavior, of a Wizard-of-Oz conversational agent. Following the experience, participants are surveyed about their intention to use the agent, their desire to cooperate with the agent, and the agent's usability. Contrary to the current tendency of designers to use high competence metaphors to describe AI products, we find that metaphors that signal low competence lead to better evaluations of the agent than metaphors that signal high competence. This effect persists despite both high and low competence agents featuring identical, human-level performance and the wizards being blind to condition. A second study confirms that intention to adopt decreases rapidly as competence projected by the metaphor increases. In a third study, we assess effects of metaphor choices on potential users' desire to try out the system and find that users are drawn to systems that project higher competence and warmth.
These results suggest that projecting competence may help attract new users, but those users may discard the agent unless it can quickly correct with a lower competence metaphor.
We close with a retrospective analysis that finds similar patterns between metaphors and user attitudes towards past conversational agents such as Xiaoice, Replika, Woebot, Mitsuku, and Tay.

\end{abstract}



\maketitle

\section{Introduction}
Collaboration between people and conversational artificial intelligence (AI) agents---AI systems that communicate through natural language~\cite{Grudin2019chatbots}---is now prevalent. As a result, there is increasing interest in designing these agents and studying how users interact with them~\cite{abokhodair2015dissecting,cranshaw2017calendar,ferrara2016rise,Grudin2019chatbots,luger2016like,shamekhi2018face}.
While the technical underpinnings of these systems continue to improve, we still lack fundamental understanding of the mechanisms that influence our experience of them. What mechanisms cause some conversational AI agents to succeed at their goals, while others are discarded?
Why would Xiaoice~\cite{shum2018eliza} amass millions of monthly users, while the same techniques powering Tay~\cite{hunt2016tay} led to the agent being discontinued for eliciting anti-social troll interactions? Many AI agents have received polarized receptions despite offering very similar functionality: for example, Woebot~\cite{nutt2017woebot} and Replika~\cite{replika} continue to evoke positive user behavior, while Mitsuku~\cite{worswick2015mitsuku} is often subjected to dehumanization. Even with millions of similar AI systems available online~\cite{chang2018conversational,lee2018chatots}, only a handful are not abandoned~\cite{Grudin2019chatbots,Zamora:2017:ISD:3125739.3125766}. The emergence of social robots and human-AI collaborations has driven home a need to understand the mechanisms that inform users' evaluations of such systems. 

In HCI, experiences of a system are typically understood as being mediated by a person's mental model of that system~\cite{norman1988psychology}.
Conveying an effective understanding of the system's behavior can enable users to build mental models that increase their desire to cooperate with the system~\cite{kocielnik2019will,bansal2019beyond,Cassell:2001:HCS:371552.371555,jakesch2019ai}. However, a mental model explanation is insufficient to answer the present question: in the case of Xiaoice and Tay, both agents were based on the same underlying technology from Microsoft, but they resulted in very different reactions by users. Likewise, other agents such as Replika and Mitsuku elicit very different evaluations while existing even within the same cultural context. While theories of mental models and culture each help us understand how users experience conversational AI agents, we require additional theoretical scaffolding to understand the phenomenon.

\begin{figure}[t]
    \centering
    \includegraphics[width=0.8\linewidth]{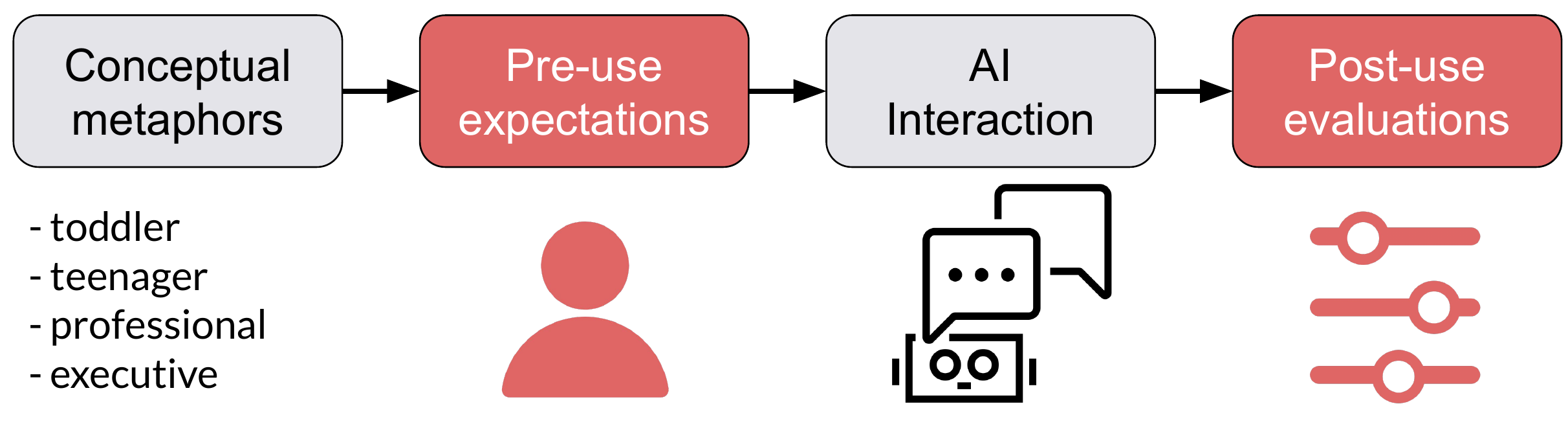}
    \caption{We explore how the metaphors used to describe an AI agent, by influencing pre-use expectations, have a downstream impact on evaluations of those AI agents.}
    \label{fig:pull_figure}
\end{figure}

An important and unexamined difference between these otherwise similar agents are the different \textit{metaphors} that they project.
Conceptual metaphors are short descriptions attached to a system that are suggestive of its functionality and intentions~\cite{mcglone1996conceptual,crawford2009conceptual}. For instance, Microsoft described Tay as an ``AI that's got no chill''~\cite{Sandvig}, while it markets Xiaoice as an ``empathetic ear''---two very different metaphors. Metaphors are a central mechanism in the designer's toolkit.
Unlike mental models, they offer more than just functional understandings of the system---they shape users' expectations from the system. 
And while most existing expectation-shaping mechanisms depend on the functionality of the specific AI system or task~\cite{kocielnik2019will}, metaphors are agnostic to specificities of a system and can be used to shape expectations for nearly any AI system.
Prior theory suggests that pre-use expectations of AI systems influence both initial behaviors~\cite{hartmann2008framing,klaaren1994role,wilson1989preferences} and long-term behaviors~\cite{kujala2017role}, even if the system itself remains unchanged while varying user expectations~\cite{raita2011too}.

We propose that these metaphors are a powerful mechanism to shape expectations and mediate experiences of AI systems. If, for example, the metaphor primes people to expect an AI that is highly competent and capable of understanding complex commands, they will evaluate the same interaction with the system differently than if users expect their AI to be less competent and only comprehend simple commands (Figure~\ref{fig:pull_figure}). Similarly, if users expect a warm, welcoming experience, they will evaluate an AI agent differently than if they expect a colder, professional experience --- even if the interaction with the agent is identical in both cases.

In this paper, we test the effect of metaphors on evaluations of AI agents. We draw on the Stereotype Content Model (SCM) from psychology ~\cite{fiske2018model,cuddy2008warmth}, which demonstrates that the two dimensions of \textit{warmth} and \textit{competence} are the principal axes of human social perception. Judgements along these dimensions provoke systematic cognitive, emotional, and behavioral reactions~\cite{cuddy2008warmth}. The SCM suggests that user expectations and therefore evaluations, are mediated by judgements of warmth and competence. We crowdsource the labeling of a set of metaphors along these axes to identify a set of metaphors that appear in different quadrants of the SCM --- e.g., a toddler, who is high warmth and low competence, and a shrewd executive, who is low warmth and high competence.

We perform an experiment ($N=260$) that manipulates the metaphor associated with an AI agent and measures how it invokes expectations of competence and warmth and how those two dimensions affect ratings of usability, intention to adopt, and desire to cooperate. We draw on an established method from prior experiments~\cite{ho2018psychological,wallis2005trouble,thies2017you,DBLP:journals/corr/abs-1904-02743} to instantiate the agent itself as a remote Wizard-of-Oz who is blind to the condition and randomized across conditions for each participant. Participants are first exposed to the agent's metaphor, then converse with the agent to complete a travel planning task~\cite{asri2017frames}.

Our results suggest that, contrary to how designers typically describe their AI agents, low competence metaphors lead to increases in perceived usability, intention to adopt, and desire to cooperate relative to high competence metaphors. These results persist despite both the low competence and high competence agents operating at full human-level performance levels via a wizard, suggesting that no matter how competent the agent actually is, people will view it negatively if it projects a high level of competence. Participants perceive the wizards to possess lower competence than the expectations implied by high competence metaphors. These results align with Contrast Theory~\cite{sherif1958assimilation}, which states that users' evaluations are defined by the difference between their experiences and expectations. Finally, we find that the warmth axis operates conversely to competence: users viewed the AI with higher warmth more positively, interacted with it longer, and were more willing to cooperate with it. This result aligns with Assimilation Theory~\cite{sherif1958assimilation}: users recolor warmth experiences in light of their initial expectation.

Previous work has sought explanations for user behavior and evaluations of AI by profiling users~\cite{dechurch2010cognitive,bansal2019beyond} or by making the AI more interpretable~\cite{caruana2015intelligible,rudin2018please,lage2018human}. However, these approaches fail to explain why otherwise functionally similar systems elicited vastly different user responses.
Our analysis suggests that designers should carefully analyze the effects of metaphors that they associate with the AI systems they create, especially whether they are communicating expectations of high competence.
In discussion, we consider implications for design by retrospectively analyzing the metaphors used to describe existing and past AI agents, such as Xiaoice, Tay, and Mitzuku, and show that our results are consistent with the adoption and user cooperation with these products. The connection between our conclusions and the outcomes experienced by Xiaoice and Tay cannot explain the whole story; however, the pattern is striking and motivates the need for exploration of mechanisms to shape expectations and elicit prosocial user behavior.

We begin by laying out related work, deriving our research question and hypotheses from prior theories. We then describe our procedure for sampling metaphors. In Study 1, we study the effects of metaphor warmth and competence. In Study 2, we sample additional metaphors along the competence axis in order to understand the effects of competence at a more fine-grained level. In Study 3, we test the negative effects of portraying a low competence metaphor by studying the effect that warmth and competence have on participants' interest in using the system in the first place. Finally, we discuss the implications of our findings for the choice of metaphors when designers deal with the dual objective of attracting more users and ensuring a positive user experience.
\section{Related Work}

Pre-use expectations play a critical role in users' initial usage of a system or design~\cite{Hartmann:2008:FUE:1357054.1357190,klaaren1994role,wilson1989preferences}. Setting positive or negative expectations colors users' evaluation of what would otherwise be identical experiences~\cite{raita2011too}. The effects of these pre-use expectations can have effects on evaluations even after weeks of interaction with a service~\cite{kujala2017role}. 

In the case of AI systems, which are often data-driven and probabilistic, there exists no simple method of setting user expectations. Providing users with performance metrics does not establish an accurate expectation for how the system behaves~\cite{kocielnik2019will}. In the absence of effective mental models of AI systems, users instead develop folk theories --- intuitive, informal theories --- as expansive guiding beliefs about the system and its goals~\cite{gelman2011concepts,sease2008metaphor,lakoff2008metaphors,french2017s}.

Prior work has shown how subjective evaluations of interface agents are strongly influenced by the face, voice, and other design aspects of the agent~\cite{xiao2004empirical,nass2005wired}, beyond just the actual capabilities of the agent. These results motivate our study of how metaphors set expectations that affect how users view and interact with conversational AI systems. Inaccurate expectations can be consequential. Previously, interviews have established that expectations from conversational agents such as Siri, Google Assistant, and Alexa are out of sync with the actual capabilities and performance of the systems~\cite{luger2016like, Zamora:2017:ISD:3125739.3125766}. So, after repeatedly hitting the agent's capability limits, users retreat to using the agents only for menial, low-level tasks~\cite{luger2016like}. While these prior interview-based studies have demonstrated that a mismatch between user expectations and system operation are detrimental to user experiences~\cite{luger2016like}, they haven't been able to establish causality and quantify the magnitude of this effect. This gap motivates our inquiry into understanding mechanisms that might shape these expectations and measuring the effect of expectations on user experiences and attitudes. We are guided by the following research question:

\begin{quote}
\textsc{Research Question}: \textit{How do metaphors impact evaluations of interactions with conversational AI systems?}
\end{quote}

\subsection{Metaphors shape expectations}
Conceptual metaphors are one of the most common and powerful means that a designer has to influence user expectations. We refer to a conceptual metaphor (or user interface metaphor, or just metaphor) as the  understanding and expression of complex or abstract ideas using simple terms~\cite{lakoff2008metaphors}. Metaphors are attached to all types of AI systems, both by designers to communicate aspects of the system and by users to express their understanding of the system. For instance, Google describes its search algorithm as a ``robotic nose''~\cite{french2017s} and YouTube users think of the recommendation algorithm as a ``drug dealer''~\cite{wu2019agent}. Starting with the desktop metaphor for personal computing in the Xerox Star~\cite{kimball1982designing}, conceptual metaphors proliferated through the design of user interfaces --- trash cans for deleted files, notepads for freetext notes, analog shutter clicking sounds for mobile phone cameras, and more. 

Some AI agents utilize metaphors based in personas or human roles, for example an administrative assistant, a teenager, a friend, or a psychotherapist, and some are metaphors grounded in other contexts, for example a Jetsons-style humanoid servant robot. Such metaphors are meant to help human-AI collaboration in complex domains by aiding users' ability to understand and predict the agent's behavior~\cite{dechurch2010cognitive,bansal2019beyond}. 
Metaphors include system descriptions outside of those rooted in human roles as well: Google describing its search algorithm as a ``robotic nose''~\cite{french2017s} and Microsoft's Zo marketed as a bot that ``Will make you LOL''. The notion of ``metaphors'' extends beyond conversational AI to non-anthropomorphic systems that ``personas'' or ``roles'' may be ill-equipped to describe.
Metaphors are effective: they influence a person's folk theories of an AI system even before they use it~\cite{DeVito:2018:PFF:3173574.3173694}.
Prior work has developed methods to extract conceptual metaphors~\cite{sease2008metaphor,lakoff2008metaphors} for how people understand AI systems and aggregate them into underlying folk theories~\cite{french2017s}.

Metaphors impact expectations, sometimes implicitly by activating different norms, biases, and expectations. For example, social robots that are racialized as Black or Asian are more likely to be subject to antisocial behaviour such as aggression and objectification~\cite{strait2018robots}. Similarly, female-gendered robots can elicit higher levels of dehumanisation than male-gendered bots. Antisocial behavior leads to verbal disinhibition toward AI systems~\cite{strait2018verbal}, and in some extreme cases, to physical abuse and even dismemberment~\cite{salvini2010safe,brscic2015escaping}.
Female voice agents are viewed as friendlier but less intelligent~\cite{nass2005wired}. Users also have a higher tendency to disclose information to female gendered agents~\cite{nass2005wired}. Race and gender of pedagogical agents affect learning outcomes---agents racialized as Black or female-gendered lead to improved attention and learning~\cite{baylor2004pedagogical}. Beyond race and gender, agents portrayed as less intelligent, taking on roles such as ``motivator'' or ``mentor'', promote more self-efficacy than agents projected as ``experts''~\cite{baylor2004pedagogical}. Young, urban users respond positively to bots that can add value to their life by suggesting recommendations, while in the role of a ``friend''~\cite{thies2017you}.

However, designers typically aim to use metaphors to affect expectations in more explicit, controlled, and pro-social ways. Most obviously, a metaphor communicates expectations of what can and cannot be done with an AI agent~\cite{kimball1982designing}. Just as we expect an administrative assistant to know our calendar but not to know the recipe for the best stoat sandwiches, an AI agent that communicates a metaphor as an ``administrative assistant'' projects the same skills and boundaries. In a similar vein, describing an agent as a ``toddler'' suggests that the agent can interact in natural language and understand some, but not all, of our communication.

While other expectation shaping mechanisms for AI agents such as tutorials and instructions have been studied~\cite{kocielnik2019will}, the effect of metaphors on user expectations and evaluations have not.
Our work also bridges to research suggesting that people already form metaphor-based theories of socio-technical systems~\cite{french2017s} and suggests design implications for how designers should choose their metaphors.

\subsection{Competing predictions: assimilation vs. contrast}
As people view AI agents as social agents~\cite{reeves1996media}, the metaphor---and thus the nature of that agent---is likely to influence their experience. However, the literature presents two competing theories for how changes to the metaphor --- and thus to expectations --- will impact user evaluation of an AI system. Assimilation theory~\cite{sherif1958assimilation} states that people adapt their perceptions to match their expectations, and thus adjust their evaluations to be positively correlated with their initial expectations. (As Dumbledore points out to Snape in Harry Potter and the Deathly Hallows, ``You see what you expect to see, Severus.'') Assimilation theory argues that users don't perceive a difference between their pre-use expectations and actual experiences. Prior work supports that, for interactive systems, users' expectations do influence evaluations~\cite{hartmann2008framing,van2008modelling}. For example, users rate an interactive system higher when they are shown a positive review of that system before using it, and rate the system lower if they are shown a negative review before using it~\cite{raita2011too}. Likewise, humor and other human-like characteristics that create high social intelligence expectations can be crucial in producing positive evaluations~\cite{liao2018all,jain2018evaluating}.

Assimilation theory would predict that a metaphor signaling high competence will set positive expectations and subsequently lead to positive evaluation:

\begin{hyp}[H\ref*{hyp:assimilation}] \label{hyp:assimilation} Positive metaphors (e.g., high competence, high warmth) will lead to higher average intention to adopt and desire to cooperate with an AI agent than if it had no metaphor or negative metaphors.
\end{hyp}

Contrast theory~\cite{sherif1958assimilation}, on the other hand, attributes user evaluations to the difference they perceive between expectations and actual experience. Contrast theory argues that we are attuned not to absolute experiences, but to differences between our expectations and our experiences. For example, exceeding expectations results in high satisfaction, whereas falling short of expectations results in lower satisfaction. This suggests that it is beneficial to set users' initial expectations to be low (with practitioners reasoning in the manner of George Weasley, in Harry Potter and the Order of the Phoenix, ```E' is for `Exceeds Expectations' and I've always thought Fred and I should've got `E' in everything, because we exceeded expectations just by turning up for the exams.'')
Users of conversational AI agents such as Alexa stumble onto humorous easter egg commands that raise their expectations of what the system can do, but then report disappointment in the contrast to discovering the system's actual limits~\cite{luger2016like}. Likewise, ratings of interactive games are driven in part by contrasting players experiences against their expectations of the game~\cite{michalco2015exploration}.

Contrast theory predicts that positive metaphors will backfire because AI agents inevitably make mistakes and have limits:

\begin{hyp}[H\ref*{hyp:contrast}] \label{hyp:contrast} Positive metaphors (e.g., high competence, high warmth) will lead to lower average intention to adopt and desire to cooperate with an AI agent than if it had no metaphor or negative metaphors.
\end{hyp}

\section{Methods}
Our research aim is to study the effect of metaphors on experiences with AI agents. So, we seek an experimental setup where participants accomplish a task in collaboration with an AI system, while avoiding effects introduced by idiosyncrasies of any particular AI system. We situate our method in goal-oriented conversational agents (or task-focused bots) as these systems represent a broad class of agents in research and product~\cite{nutt2017woebot,shum2018eliza,worswick2015mitsuku,hunt2016tay,woollaston2016following,damani2018ruuh}.

\subsection{Collaborative AI task}
Goal-oriented AI systems, such as those for booking flights, hotel rooms, or navigating customer service requests, have become pervasive on social media platforms including Kik, Slack, and Facebook Messenger, with as many as one million flooding the Web between $2015$ and $2019$~\cite{Grudin2019chatbots}. Surveys revealed that as of 2018, as many as 60\% of surveyed millenials had used a chatbot~\cite{Arnold} and 15\% of surveyed internet users had used customer-service chatbots~\cite{salesforce}. Their prevalence means that interaction with such an agent is an ecologically valid task, and that many users online are familiar with how to interact with them. We draw on a common set of transactional tasks such as appointment booking, scheduling, and purchasing, which require people to engage with the agent in task-focused dialogue to acquire information or complete their task~\cite{DBLP:journals/corr/abs-1809-08267}. Inspired by the popular Maluuba Frames~\cite{asri2017frames} data collection task templates, used to evaluate conversational agents in the natural language processing community, we utilize a travel planning task. More concretely, the task is a vacation planning endeavor where users must pick a vacation package that meets a set of experimenter-specified requirements through a search-compare-decide process. Specifically, every participant is presented with the following prompt:
\begin{quote}
You are considering going to New York, Berlin or Paris from Montreal. You want to travel sometime between August 23rd and September 1st.  You are traveling alone. Ask for information about options available in all cities. Compare the alternatives and make your decision.
\end{quote}
Participants were further instructed to determine what they could get for their money and to take into consideration factors they would consider while actually planning a vacation, including wifi, breakfast options, and a spa. The task is structured to involve three sub-goals: finalize a hotel package, an outgoing flight and an incoming flight back to Montreal. 

\subsection{Wizard-of-Oz conversational agent}

We sought a conversational AI agent whose actual performance was strong enough for our result to generalize as the underlying AI models improve. So, following a pattern in prior work~\cite{ho2018psychological,wallis2005trouble,thies2017you,DBLP:journals/corr/abs-1904-02743}, we adopt a Wizard-of-Oz study paradigm. 

We hire and train customer-support professionals from the Upwork platform to act as wizards in our experiment and pay them their posted hourly rate of $\$10-\$20$ per hour. The wizards play the role of the conversational AI agent in the text chat. We filtered workers who had at least a $90\%$ job success rating from past work, and had already earned $\$20k$ USD through the Upwork platform. We also filtered for workers with English proficiency by asking them to submit a cover letter detailing their past work experience and manually checked for spelling or grammatical errors. We hired 5 wizards in all.
To eliminate wizard-specific confounds across our different conditions, wizards were blind to the treatment condition of the participants in the study, and randomized to a new condition for each new participant that they interacted with. Randomizing this source of variation produces an unbiased estimate of the effect of each condition.

Wizards were trained on how to provide responses by engaging in practice tasks with the authors as participants. They were instructed to send emotionally neutral responses simply addressing the participant's query. Wizards were required to, at every turn, search for hotels that met the requirements specified by the participant. In case multiple hotels met the set of requirements, they were asked to present all the options to the participant, ordered according to how they appear in the database. Once users achieved their three goals (booking two flights and a hotel), the wizards informed the users to proceed to the next step of the study. In order to minimize inter-wizard differences, we provided feedback to wizards from trial conversations to ensure that there were no drastic differences across the wizards’ performance. In many common exchanges, wizards were also provided with template responses. To highlight the similarity of the wizards’ responses, sample responses of three wizards to similar input queries are provided in the Supplementary Material.

Task-focused agents are knowledgeable within a narrow task focus~\cite{Grudin2019chatbots} and are often unable to answer questions that require external knowledge.
In order to retain phenomena associated with access to finite knowledge, we provide our wizards with a database of hotels and flights. We construct this database by hand to mimic what one would find on a standard travel booking platform. We have provided details on the construction of the database and examples of hotels and flights from the database in the supplementary material. Consistent with instructions provided to wizards in the creation of Frames~\cite{asri2017frames} corpus, if the wizard is asked about knowledge that is outside of their available database, wizards were trained to respond that they do not have that information.

We constructed a conversational chat platform via using Chatplat (\url{https://www.chatplat.com}). We embedded this chat widget into our web based survey.

\subsection{Sampling metaphors}
We place participants into treatment groups each defined by the metaphor used to describe their AI collaborator. Instead of randomly sampling metaphors, we draw on the Stereotype Content Model (SCM)~\cite{fiske2018model,cuddy2008warmth}, an influential psychological theory that articulates two major axes in social perception: warmth and competence. These two dimensions have proven to far outweigh others and repeatedly come up as prime factors in literature~\cite{asch1946forming,judd2005fundamental,wojciszke2005affective}. Judgements on warmth and competence are made within $100$ milliseconds~\cite{todorov2008understanding} and a change in these traits alone can wholly change impressions~\cite{zanna1972attribute}. The SCM proposes a quadrant structure and cognitive notions of warmth and competence- better understood as discrete- are characterized as being low or high~\cite{cuddy2008warmth}.  
Warmth is characterised by notions such as good-naturedness and sincerity, while competence is characterised by notions of intelligence, responsibility, and skillfulness. For example, a ``shrewd travel executive'' can be described as high competence and low warmth.
We sample metaphors such that they have either high or low values of warmth and competence.

\begin{figure}[tb]
    \centering
    \includegraphics[width=0.8\linewidth]{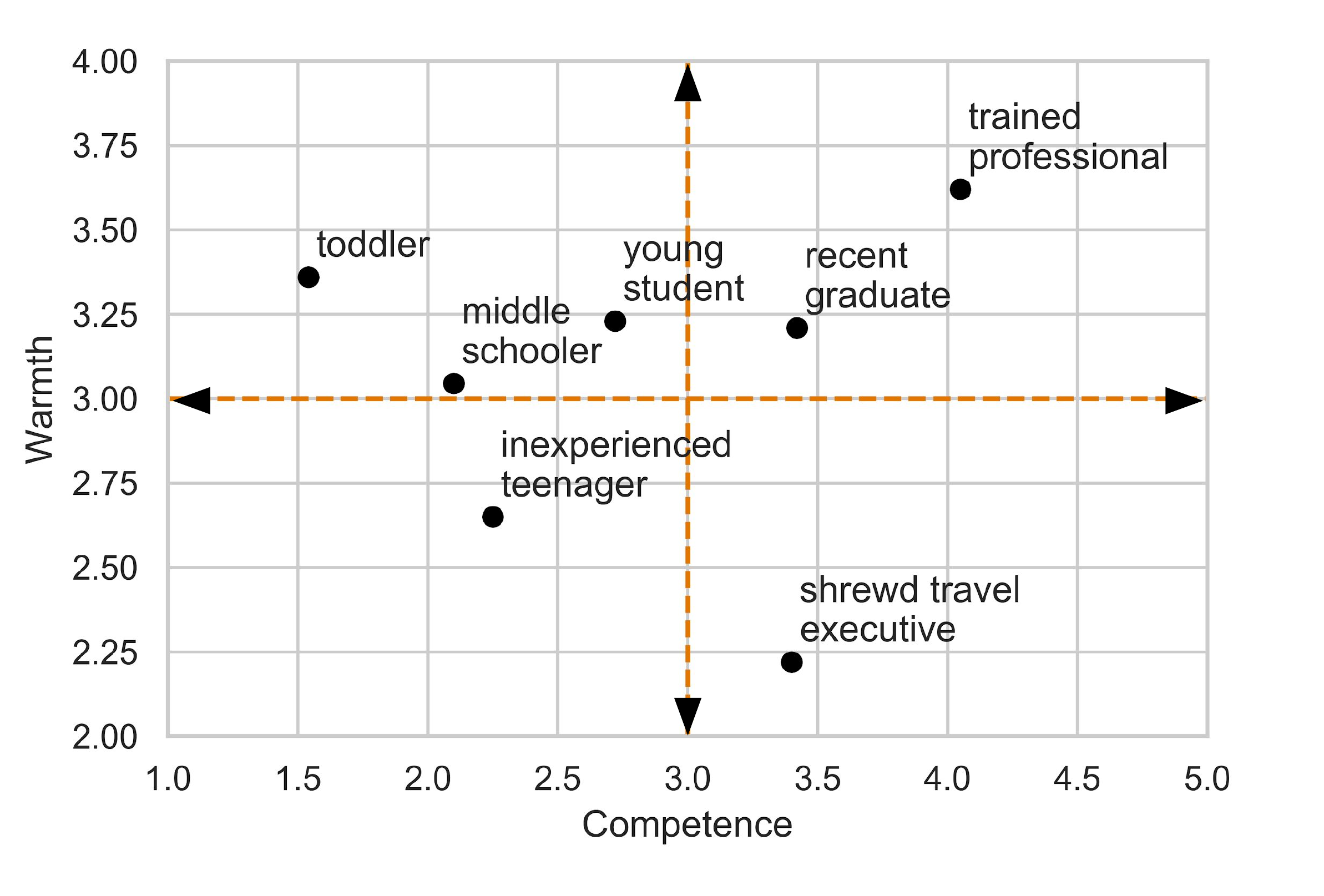}
    \caption{Average warmth and competence measured for the conceptual metaphors sampled for our studies. Both the axes ranged from $1$ to $5$.}
    \label{fig:metaphors}
\end{figure}

\begin{footnotesize}
\begin{table}[tb]
\begin{tabular}{llllllll}
           & Toddler 
           & \begin{tabular}[c]{@{}l@{}}Middle \\ Schooler\end{tabular}
           & \begin{tabular}[c]{@{}l@{}}Inexperienced \\ Teenager\end{tabular} 
           & \begin{tabular}[c]{@{}l@{}}Young \\ Student\end{tabular}
           & \begin{tabular}[c]{@{}l@{}}Shrewd Travel \\ Executive\end{tabular}  
           & \begin{tabular}[c]{@{}l@{}}Recent \\ Graduate\end{tabular}
           & \begin{tabular}[c]{@{}l@{}}Trained \\ Professional\end{tabular} \\ \hline
Competence & \begin{tabular}[c]{@{}l@{}}$1.54 \pm 1.04$\end{tabular}  
           & \begin{tabular}[c]{@{}l@{}}$2.1 \pm 0.83$\end{tabular} 
           & \begin{tabular}[c]{@{}l@{}}$2.25 \pm 0.85$\end{tabular}
           & \begin{tabular}[c]{@{}l@{}}$2.72 \pm 0.58$\end{tabular} 
           & \begin{tabular}[c]{@{}l@{}}$\textbf{3.4} \pm 0.97$\end{tabular}
           & \begin{tabular}[c]{@{}l@{}}$\textbf{3.42} \pm 0.72$\end{tabular} 
           & \begin{tabular}[c]{@{}l@{}}$\textbf{4.05} \pm 0.58$\end{tabular} \\ 
Warmth     & \begin{tabular}[c]{@{}l@{}}$\textbf{3.36} \pm 0.81$\end{tabular}
           & \begin{tabular}[c]{@{}l@{}}$\textbf{3.05} \pm 0.69$\end{tabular} 
           & \begin{tabular}[c]{@{}l@{}}$2.65 \pm 0.73$\end{tabular}
           & \begin{tabular}[c]{@{}l@{}}$\textbf{3.23} \pm 0.56$\end{tabular} 
           & \begin{tabular}[c]{@{}l@{}}$2.22 \pm 1.02$\end{tabular}
           & \begin{tabular}[c]{@{}l@{}}$\textbf{3.21} \pm 0.67$\end{tabular} 
           & \begin{tabular}[c]{@{}l@{}}$\textbf{3.62} \pm 0.59$\end{tabular}     
\end{tabular}
\caption{Warmth and competence values (average $\pm$ standard deviations) for the metaphors we use across the studies. High competence and warmth values are in bold.}
\label{tab:metaphor_stds}
\end{table}
\end{footnotesize}

In our first study, we use four metaphors, one in each quadrant, to study the impact of competence and warmth. We pre-tested a set of metaphors for a conversational agent, measuring the perceived competence and warmth of conversational AI agents described with these metaphors using a $5$ point Likert scale. We captured $50$ ratings for each metaphor from workers on Amazon Mechanical Turk --- a mutually exclusive set of workers from those who will later be involved in the experiment. Based on the results (see Figure~\ref{fig:metaphors}), we chose  ``trained professional travel assistant'' (high competence, high warmth), ``shrewd travel executive'' (high competence, low warmth), ``toddler'' (low competence, high warmth), and ``inexperienced teenager'' (low competence, low warmth). We selected metaphors that were otherwise agendered, with similar socio-cultural connotations across the world, and representative of actual metaphors that could be associated with a travel assistant bot. These four metaphors form our four treatment groups in Study 1; their mean and standard deviation values of competence and warmth are reported in Table~\ref{tab:metaphor_stds}.

In Study 2, we follow the same procedure to characterize several additional metaphors: ``middle schooler'', ``young student'', and ``recent graduate''. These metaphors offer intermediate levels of competence, with ``toddler'' less competent than ``middle schooler'', ``middle schooler'' less competent than ``young student'', and ``young student'' less competent than ``trained professional''. ``Young student'' is associated with higher competence levels than ``middle schooler'', suggesting that people's impression of a ``young student'' is a high schooler or college student, somewhere between a ``middle schooler'' and a ``recent graduate''. In Study 3, we revisit the metaphors we analyzed in Study 1 to understand the effects of metaphors on potential users' likelihood of trying out the system and their intentions of co-operating with it prior to using it.


\section{Study 1: Metaphors drive evaluations}
In our first study, we examine the effects of metaphors attached to the conversational system on user pre-use expectations and post-use evaluations. Specifically, we examine participants' perceived pre- and post-use usability and warmth of the AI system. Additionally, we measure their post-use intention to adopt and their desire to cooperate with such a system given their treatment metaphor condition. Finally, we analyze the chat logs to explore if there are behavioral differences between the participants in different conditions.

\subsection{Procedure}
We perform a between-subjects experiment where participants in each treatment condition are primed with a metaphor to associate with the system. As described in the previous section, metaphors were chosen to vary as low/high warmth $\times$ low/high competence, resulting in four treatment conditions. In addition, we included a control condition where participants were not primed with a metaphor, resulting in five total conditions.

After consenting to the study, participants were introduced to one of the study conditions, i.e.~they were shown one of the four metaphors, or a control condition of no metaphor:
\begin{quote}
    \textit{The bot you are about to interact with is modeled after a ``shrewd travel executive''.}
\end{quote}
With the study condition revealed, participants were asked questions about their pre-use expectations of the AI system's competence and warmth. Next, participants were shown the goal-oriented task description and allowed to interact with the wizard posing as conversational agent via the chat widget until they completed their task. 

After finalizing their travel plans, participants were asked to evaluate their experience with the AI system and answer the manipulation check question. Finally, participants were debriefed, informed of the actual purpose of the study, and made aware that they were talking to a human and not an AI system. A high-level workflow is depicted in Figure~\ref{fig:pull_figure}.

\subsection{Measures}

\noindent\textbf{User evaluation measures.}
To test contrast theory, it is important to measure a user's evaluation of the experience without drawing explicit attention to the contrast between their expectations and their experience---since this makes the contrast salient~\cite{kujala2017role}. 
So, we independently measure pre-use expectations and post-use evaluations without explicitly asking if expectations were met or violated. To gauge participants' pre-use expectations and post-use perceptions of the systems competence and warmth, we ask the participants to report how strongly, on a $5$ point Likert scale (where 1 = strongly disagree and 5 = strongly agree), they agree with the following statements, both before and after they interacted with the AI system. Questions asked before use simply replaced the past tense of the verb with the future tense; the question ordering was randomized.
\begin{itemize}
    \item Usability: Since our notion of a system's competence is akin to the notion of usability in previous studies, we adapt questions from previous surveys that examine usability~\cite{kujala2017role}. These questions are: 1) \textit{``Using the AI system was (will be) a frustrating experience.''} 2) \textit{``The AI system was (will be) easy to use.''} 3) \textit{``I spent (will spend) too much time correcting things with this AI system.''} 4) \textit{```The AI system met (will meet) my requirements.''} Responses from before using the system are combined to form a pre-use usability index ($\alpha = 0.91$) while responses from after the conversation are combined to form a post-use usability index ($\alpha = 0.87$).
    \item Warmth: To measure the warmth of the AI system, we draw on different warmth levels articulated in the stereotype content model~\cite{doi:10.1111/0022-4537.00128}: 1) \textit{``This AI system was (will be) good-natured.''} 2) \textit{``This AI system was (will be) warm.''} Responses from before using the system are combined to form a pre-use warmth index ($\alpha = 0.94$). Similarly, responses from after the conversation are combined to form a post-use warmth index ($\alpha = 0.91$).
    \item Intention to Adopt and Desire to Cooperate: We borrow from prior work~\cite{kujala2017role} that captures user evaluations through their intentions to adopt the system. Since we increasingly have situations in which humans work alongside AI systems where these systems augment human efforts, it also becomes necessary to understand users' behavioural tendencies towards these systems. So, we draw on prior work in HRI~\cite{8673307} and capture users' behavioural tendencies through their desire to help and cooperate with the AI system. 
    After their interaction with the system, participants are probed for their intentions to adopt as well as their desire to cooperate with the system. To probe for the participants' intentions to adopt, we asked them the following two questions on $5$ point Likert scales: 1) \textit{``Based on your experience, how willing are you to continue using the service?''}, 2) \textit{``How likely is it that you will be using the service in the future?''}. Like previous work~\cite{kujala2017role}, these two questions are combined to form an intention to adopt index ($\alpha = 0.98$). 
    To understand the participants' desire to cooperate, we use questions about behavioral tendencies towards stereotyped groups adapted to the context of social robots~\cite{8673307}. Users are asked on a 5 point Likert scale: \textit{``How likely would you be to cooperate with this AI?''}  and \textit{``How likely would you be to help this AI?''}. Like previous work~\cite{8673307}, these two questions are combined to form a cooperation index ($\alpha = 0.79$).
\end{itemize}

\noindent\textbf{Conversational behavior measures.}
To investigate if participant behavior changes across conditions we include measures to analyze differences in the conversational behavior of users.
\begin{itemize}
    \item Language measures: To measure differences in the chatlogs by the participant and by the wizards across the various conditions, we utilize the popular linguistic dictionary Linguistic Inquiry and Word Count, known as LIWC~\cite{pennebaker2001linguistic}. LIWC uses $82$ dimensions to determine if a text uses positive or negative emotions, self-references, and causal words, to help assess physical and mental health, intentions and expectations of the writers. We categorize all the words used by the participants and the wizards into LIWC categories and create normalized frequency histograms of these categories. We compare the words used by the participants across all the conditions to see if there significant differences in the types of LIWC categories used. Similarly, we compare the wizards' words across all conditions. Finally, we combine the words used by the wizard and participant together and also check to see if there were differences between conversation across the different conditions.
    \item Conversation measures: We also investigate differences across conditions, at the level of individual messages and whole conversations in terms of \textit{number of words used} and \textit{duration of interaction}.
\end{itemize}

\subsection{Participants}
For all the studies in this paper, we hired participants to interact with our Wizard-of-Oz AI system from Amazon Mechanical Turk (AMT). Participants were all US citizens aged $18+$. Each participant was allowed to take part in the experiment only once. Participants were compensated $\$4$ for a survey lasting an average of $15 \pm 5$ minutes, for a rate of roughly $\$15$/hr in accordance with fair work standards on Mechanical Turk~\cite{whiting2019fair}. Participants' data was discarded if they failed to follow instructions, left the conversation midway or did not follow the task specifications. $50.7\%$ of our participants were female and the mean age of participants was $41\pm10.3$. 

In this specific study, for a small expected effect size of $f = 0.05$, a power analysis with a significance level of $0.05$, powered at $80\%$, a power analysis indicated that we require $25$ participants per condition, or $125$ total participants. Thirteen participants' responses were discarded because the raters had coded their WoZ manipulation check as expressing suspicion that the agent might be human. After these exclusions, we had a sample size of $140$ participants, which met the requirements from our power analysis.

\subsection{Wizard-of-Oz manipulation check}
To ensure that our study was not compromised by participants who identified that they were speaking to a wizard instead of an AI, we included a manipulation check at the end of the survey. The manipulation check gauges whether the participant was suspicious of the AI without explicitly drawing their attention to the fact that this might be the case. So, drawing on prior Wizard-of-Oz studies~\cite{Hinds}, we asked the participants how they thought the system worked from a technical standpoint. 

The responses were sent to two coders who inspected each response individually and marked all the responses that suspected a person was pretending to be an AI system. Participants who expressed suspicion that the system might be human were excluded from further analysis.
Some participants were very confident they knew how such our conversational AI could be built: \textit{I am a programmer so i understand the bot has a vocabulary of words it attempts to parse through, then it takes what it finds from the user and checks against a database to output information it thinks is relevant.}
Others talked about how \textit{Most chat bots go through ``training'' beforehand to be able to parse commonly asked questions and phrasing} or how it must be using a \textit{database full of responses}. 

Out of all our participants, both coders identified the same $13$ participants ($\kappa=1$) who failed the manipulation check by calling out the agent as a human, resulting in a suspicion level of $10.4\%$. In most of these cases, it was triggered by a wizard making a typo or taking too long to respond. One suspicious participant exclaimed, ``\textit{I'm like $90\%$ sure it's not a bot, but if it were a bot, machine learning, though I don't know exactly what THAT means}''. These $13$ participants were excluded from our analysis.

\subsection{Results: Metaphors shape pre-use expectations}
We compare the impact of setting expectations by varying the competence and warmth of the metaphors used. We perform our analysis using a pair of two-way analyses of variance (ANOVAs), where competence and warmth are two categorical independent variables, and pre-use usability and warmth are the dependent variables. We compared the impact of the conceptual metaphors used to describe our system compared to the control condition to measure if they have an impact on the participants' default expectations of conversational AI systems. So, the independent variables are categorized into high, low or control categories.

\begin{figure}[t!]
    \centering
    \includegraphics[width=0.8\textwidth]{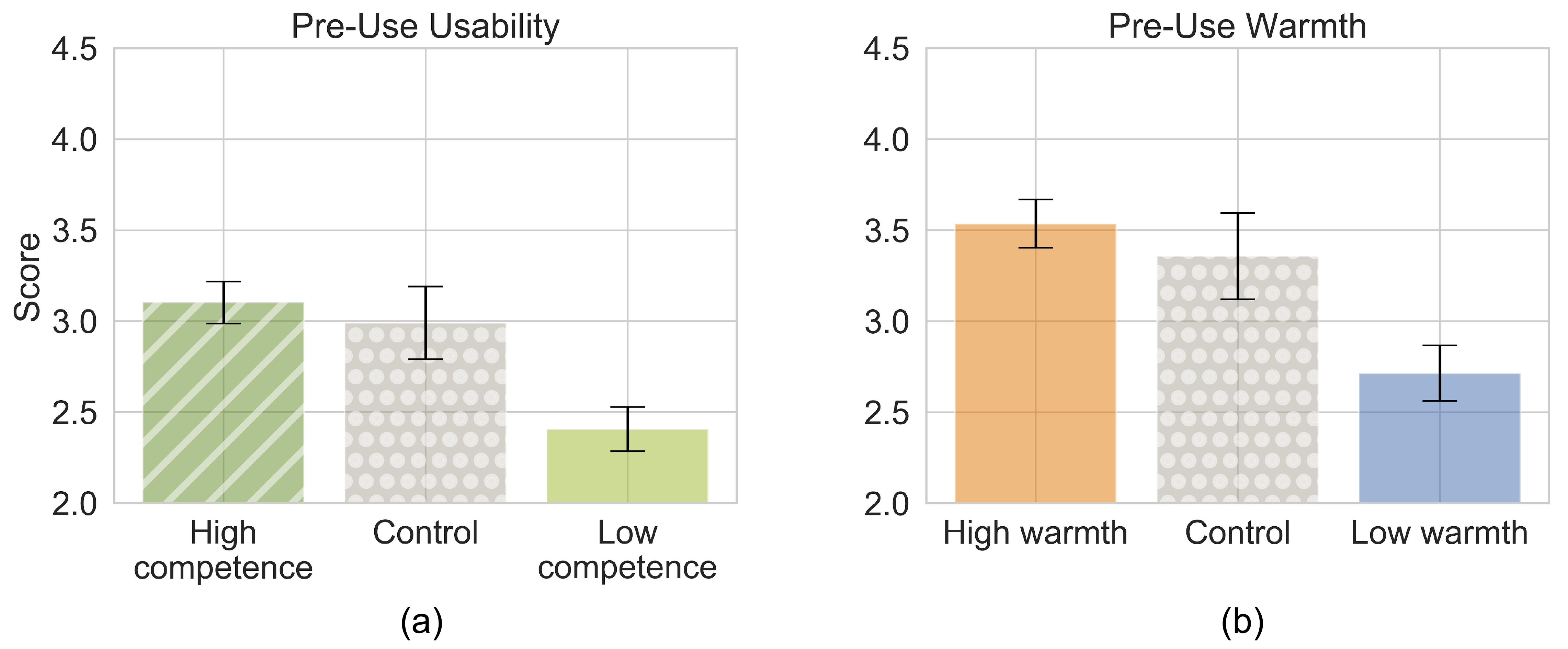}
    \caption{(a) Metaphors that signal high competence lead to higher pre-use usability scores. (b) Similarly metaphors that signal high warmth lead to higher pre-use warmth scores. We also notice from both (a) and (b) that participants are naturally predisposed to have high expectations of usability and warmth from conversational systems; however, priming them with metaphors reduces the variance of their expectations as opposed to when their expectations are uninformed.}
    \label{fig:pre-use}
\end{figure}

Pre-use usability is affected by the metaphor's competence. For pre-use usability (Figure~\ref{fig:pre-use}~(a)), we find that competence has a large main effect $(F(1, 108) = 8.71, p < 0.001, \eta^2 = 0.12)$. By default, participants have high expectations of competence. A post-hoc Tukey revealed that pre-use usability was significantly lower for the low competence condition $(2.41 \pm 0.89)$ than high competence $(3.10 \pm 0.86, p < 0.001)$ or control $(2.99\pm 1.03, p = 0.019)$ conditions. We found no main effects for warmth.

Pre-use warmth is likewise affected by the metaphor's warmth. For pre-use warmth (see Figure~\ref{fig:pre-use}~(b)), we find that warmth has a large main effect $(F(1, 108) = 8.20, p < 0.001, \eta^2 = 0.11)$. By default, participants have high expectations of warmth. A post-hoc Tukey revealed that pre-use warmth is significantly low for the low warmth condition $(2.71 \pm 1.12)$ than both the high warmth $(3.53 \pm 1.05, p < 0.001)$ and control $(3.35\pm 1.23, p = 0.035)$ conditions. We found no main effects of competence.

Together, these tests imply that participants, by default, expect conversational AI to possess high competence and high warmth. However, the change in expectation caused by low competence and low warmth implies that these conceptual metaphors do affect participants' expectations of how the AI system will perform and behave. We visualize the means and standard errors for these conditions in Figure~\ref{fig:pre-use}~(a, b).

\subsection{Results: Metaphors impact user evaluations and user attitudes}

We compare the impact of varying the competence and warmth of the metaphors on participants' post-use evaluations of the AI system's usability and warmth. We perform our analysis using a pair of two-way ANOVAs where competence and warmth are categorical independent variables, and post-use usability and post-use warmth are the two dependent variables.

\begin{figure*}[t!]
    \centering
    \includegraphics[width=0.85\linewidth]{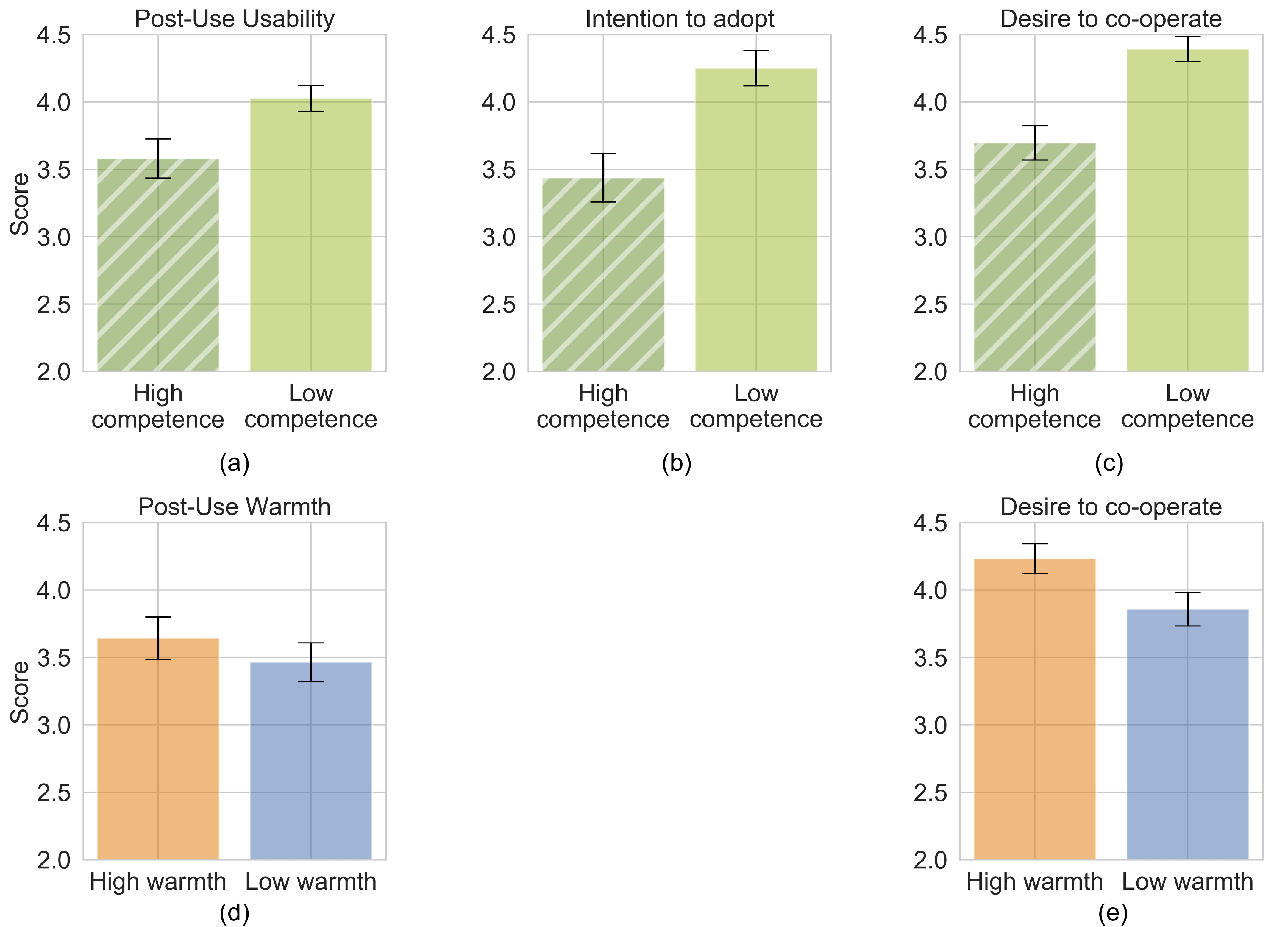}
    \caption{The low competence metaphor condition features the highest post-use usability, intention to adopt, and desire to cooperate. This result suggests that metaphors that undersell the AI agent's competence are most likely to succeed.
    }
    \label{fig:post-use}
\end{figure*}

Participants perceive agents with low competence to be more usable after interaction. For post-use usability (Figure~\ref{fig:post-use}~(a)), competence has a main effect $(F(1, 108) = 6.49, p < 0.05, \eta^2 = 0.06)$. Competence has a smaller effect on post-use usability than on pre-use usability, implying that the actual interaction of the participant with the system affects their final evaluations. In the low competence condition, post-use usability was rated at $(4.02 \pm 0.72)$ and in the high competence condition, it was rated $(3.58 \pm 1.07)$. These results suggest that users perceive a difference between their experience and their expectations in terms of competence of the agent. 
The means for post-use usability are also higher than for pre-use usability for both high $(t(55)=-3.48, p < 0.001)$ and low $(t(55)=-10.79, p < 0.001)$ competence conditions. 
We found no main effects of warmth or interaction effects between competence and warmth.

We observe post-use warmth ratings to be higher in the high warmth condition than the low warmth condition though the difference is not significant.
For post-use warmth (Figure~\ref{fig:post-use}~(d)), we find no main effects of competence or warmth and no interaction effects. In the high warmth condition, warmth was rated at $(3.64 \pm 1.17)$ and in the low warmth condition, it was rated $(3.50 \pm 1.12)$. 
There is no significant difference between the means of pre-use and post-use warmth for high warmth $(t(55)=-0.67, p = 0.501)$, but it is significantly different in the low warmth $(t(55)=-4.56, p < 0.001)$.  

Using the composites described in the study design, we measure the effect of conceptual metaphors on the participants' intention to adopt and desire to cooperate after interacting with the system.

Low competence metaphors increase participants' likelihood of adopting the AI agent.
For their intention to adopt (Figure~\ref{fig:post-use}~(b)), competence has a main effect $(F(1, 108) = 13.31, p < 0.001, \eta^2 = 0.12)$. In the high competence condition, the intention to adopt was rated at $(3.42 \pm 1.33)$ and in the low competence condition, it was rated $(4.25 \pm 0.96)$. These results support Hypothesis~\ref{hyp:contrast} as we see support for contrast theory: participants are more likely to adopt an agent that they originally expected to have low competence but outperforms that expectation. They are less forgiving of mistakes made by AI systems they expect to have high competence. We found no main effects of warmth or interaction effects between competence and warmth.

Participants prefer to cooperate with agents that have high warmth and low competence. For their desire to cooperate with the AI system, we found that both competence $(F(1, 108) = 20.58, p < 0.001, \eta^2 = 0.19)$ and warmth $(F(1, 108) = 5.96, p < 0.01, \eta^2 = 0.05)$ had main effects but no interaction effect (see Figure~\ref{fig:post-use}~(c, e)). The means increase from high $(3.69 \pm 0.93)$ to low competence of $(4.39 \pm 0.67)$. Similarly, the means decrease from high $(4.23 \pm 0.81)$ to low warmth $(3.85 \pm 0.92)$. These results provide mixed support to both Hypothesis~\ref{hyp:assimilation} and Hypothesis~\ref{hyp:contrast} as we see support for assimilation theory along the warmth dimension and contrast theory along the competence dimension. If participants are told that the AI system is high warmth, they are more likely to cooperate with it. But if the AI system is described as high competence, they are less likely to cooperate.

\begin{figure}[t]
    \centering
    \includegraphics[width=0.8\linewidth]{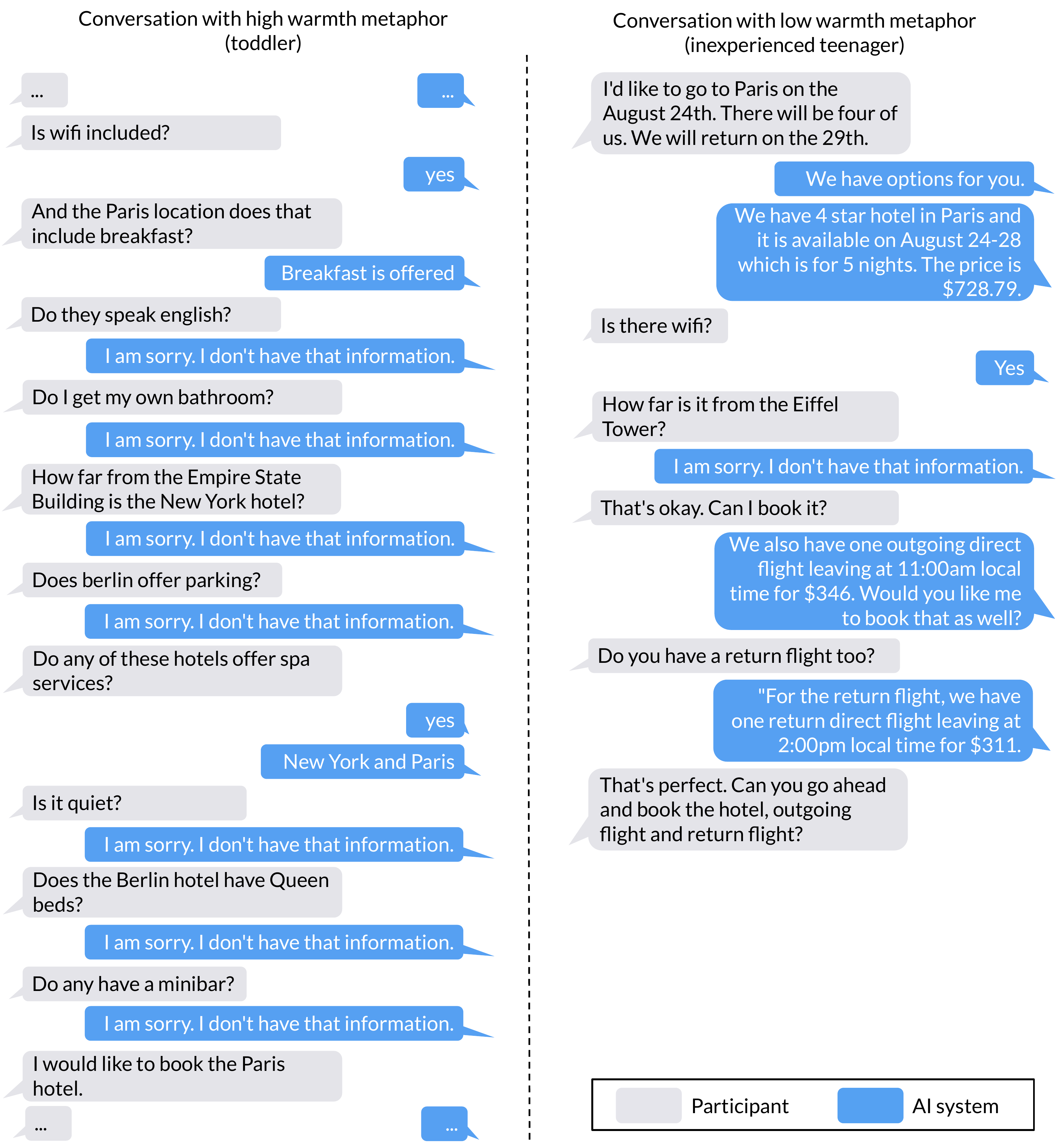}
    \caption{Segments of two example conversations between a participant with our conversational AI system. In both cases, the participant expects the AI system to have low competence. While the left conversation is in the high warmth metaphor condition, the right conversation is in the low warmth metaphor condition. Participants in the high warmth condition ask more questions and explore the space of possible interactions by asking the agent details about checked luggage and hotel amenities. Wizards, acting as conversation agents, are given a fixed knowledge set, mimicking how today's systems are designed, and reply with apologies when asked about details outside of their knowledge.}
    \label{fig:conversation_2}
\end{figure}

\subsection{Results: Expectations change, but behavior doesn't}
We analyze the chat logs with LIWC features, following a standard LIWC analysis protocol of building a frequency count of how often words belonging to a specific LIWC category were used. We contrasted these counts across the various conditions and observed no significant differences ($p>.05$) in language level phenomenon in the chatlogs across the conditions.
This result implies that the post-use evaluations are driven primarily by the expectations set by the metaphors, not by the actual content of the conversation. In other words, evaluations differed between conditions, but the actual conversations themselves did not. The wizard was blinded to the condition, so any differences would have needed to be prompted by the participant. However, we acknowledge that there might be language shifts that LIWC categories cannot capture. 

Participants use more words and spend more time speaking to agents with high warmth.
We find a significant main effect of warmth on the number of words used per conversation $(F(1, 106) = 5.35, p < 0.05, \eta^2 = 0.05)$.
The number of words increase from $(82\pm37)$ in low to $(101\pm45)$ in high warmth. 
We also find that participants in the high warmth condition typically spend an average of $4\pm1.5$ minutes longer while interacting with the AI system. On a qualitative inspection, we find that participants tend to ask more questions and spent more time exploring the AI system's capabilities.
Consider, for example, the conversation shown in Figure~\ref{fig:conversation_2}, where the participant expects the bot to have low competence and high warmth. The participant asks numerous questions to test the system's capabilities and even though it fails, they later express a high intention to adopt and cooperate with the system.

\subsection{Summary}
Our results support contrast theory (Hypothesis~\ref{hyp:contrast}) for the competence axis. Users are more tolerant of gaps in knowledge of systems with low competence but are less forgiving of high competence systems making mistakes. The intention to adopt and desire to cooperate decreases as the competence of the AI system metaphor increases.
For the warmth axis, our results provide some support for assimilation theory (Hypothesis~\ref{hyp:assimilation}): users are more likely to co-operate and interact longer with agents portraying high warmth, but we do not observe significant impact of warmth on users' intention to adopt.
\section{Study 2: The competence-adoption curve}
\begin{figure}[t!]
    \centering
    \includegraphics[width=0.8\textwidth]{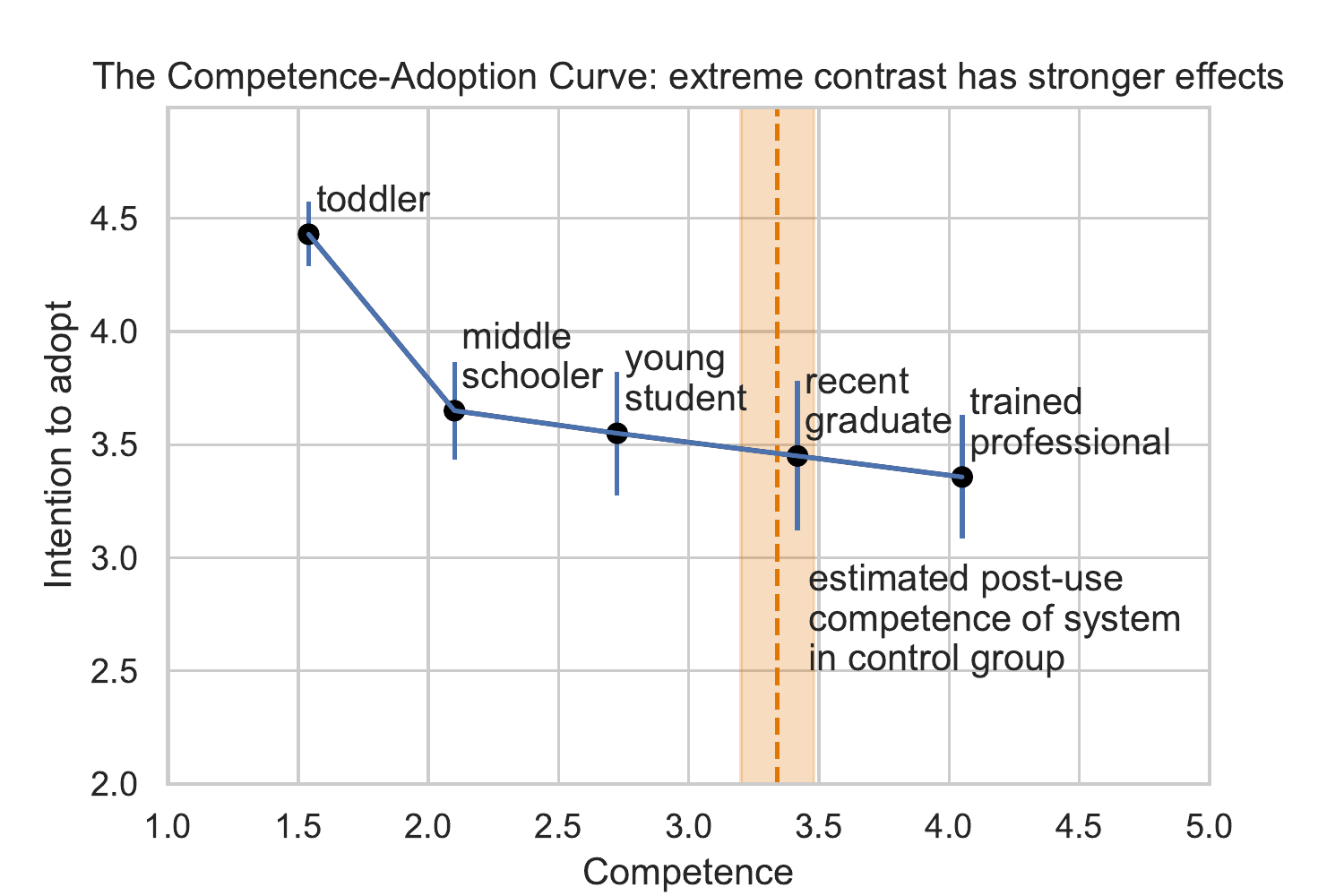}
    \caption{A larger positive violation of expectation increases adoption intentions. The intention to adopt decreases monotonically with an increase in expected competence of the system. The red vertical line shows the average score users in the control condition assigned the system and the yellow shaded region around the vertical line depicts the standard deviation.}
    \label{fig:competence}
\end{figure}

Study 1 established that setting low expectations and violating them positively increased the likelihood of users adopting the system. In Study 2, we zoom in and try to understand how user evaluations change as the magnitude of that gap changes. We sample additional metaphors and use the same experiment procedure as before to characterize how users' intentions to adopt the system vary as gap between users expectations and their experience changes. For this purpose, we rely on the same measure of Intention to Adopt as Study 1. 

\subsection{Procedure}
To precisely traverse the range of perceived competence, we sampled $3$ additional metaphors --- ``middle schooler'', ``young student'', and ``recent graduate''. Our pre-experiment survey revealed that these metaphors had perceived competence levels between the ``toddler'' and the ``trained professional''.  Together, these five metaphors formed five treatment conditions. As Figure~\ref{fig:metaphors} demonstrates, all five metaphors lie in the high warmth half of the space, minimizing any interfering effects of variations in warmth. Participants in these five conditions were primed with the respective metaphor. 

To understand users' unprimed evaluations of the system, we asked a sixth control group to participate without a metaphor (similar to the control condition in Study 1). Afterwards, we asked them to pick from the list of five metaphors and identify which one they felt described the system most accurately after use.

\subsection{Measures}
Similar to Study 1, we measure participants' intention to adopt and desire to co-operate across all five metaphors.

\subsection{Participants}
Similar to the protocol in Study 1, participants were recruited on AMT. We recruited $20$ participants for each condition, for a total of $120$ participants.
The duration of the study was similar to Study 1 and participants were compensated at the same rate. The average age of participants was $39 \pm 11.06$; $48\%$ identified as female.

\subsection{Wizard-of-oz manipulation check}
The two coders were consistent and identified the same $5$ participants ($\kappa=1$) as being suspicious, implying a low suspicion level of $4.1\%$. These five participants were removed from analysis.

\subsection{Extreme violations of expectations have stronger effects}
The five metaphors we sampled are shown in Figure~\ref{fig:competence}, where the x-axis depicts workers' perceived competence of a system with that metaphor and the y-axis depicts a different set of users' intention to adopt. The red vertical line shows the average score users in the control condition assigned the system and the yellow shaded region around the vertical line depicts the standard deviation. The unprimed system was viewed roughly as competently as a recent graduate.

Over-performing low competence leads to higher adoption than over-performing medium competence, and projecting any more competence than the toddler metaphor incurs an immediate cost (Figure~\ref{fig:competence}).
These results paint a fuller picture of contrast theory at play, as the intention to adopt decreases monotonically as the expected competence of the system increases. Consistent with prior literature, the effect is greater as the contrast is greater~\cite{geers1999affective,brown2012expectation}. However, the effect is nonlinear, with only the lowest competence metaphor receiving a substantial benefit.

The ``toddler'' metaphor sees the highest (beneficial) violation as it is furthest away from the vertical line and sees the greatest intention to adopt and desire to cooperate. There was a statistically significant difference between groups as determined by one-way ANOVA for intention to adopt $(F(4,114) = 3.701, p = .007)$ and for desire to cooperate $(F(4,114) = 6.841, p < .001)$. A Tukey post-hoc test revealed that intention to adopt was statistically significantly higher for ``toddler'' $(4.42 \pm 0.76)$ than ``young student'' $(3.55 \pm 1.25, p = .047)$, ``recent graduate'' $(3.45 \pm 1.5, p = .021)$, and ``trained professional'' $(3.35 \pm 1.43, p = .007)$. There was no statistically significant difference between other metaphors. Similarly, a Tukey post-hoc test revealed that desire to cooperate was statistically significantly higher for ``toddler'' $(4.50 \pm 0.50)$ than ``middle schooler'' $(3.52 \pm 1.08, p = .001)$, ``young student'' $(3.82 \pm 0.86, p = .006)$, ``recent graduate'' $(3.95 \pm 0.76, p = .003)$, and ``trained professional'' $(3.87 \pm 0.93, p = .013)$. There was no statistically significant difference between other metaphors.

\subsection{Summary}
Our results further support contrast theory (Hypothesis~\ref{hyp:contrast}) for the competence axis. Users are more likely to adopt a lower competence agent than one with high competence, even though all conditions were exposed to human-level performance. We additionally see an asymmetry --- users are even more likely to adopt an agent that exceeds extremely low expectations than one that exceeds slightly higher (but still low) expectations. And as the agent begins to under-perform expectations, intentions to adopt decrease further.
\section{Study 3: The cost of low-competence metaphors}
From our results so far, it might appear that that designers should pick metaphors that project lower competence and high warmth regardless of experience, as these conditions are most conducive for cooperative and patient user interactions. However, such a conclusion might be myopic. Metaphors attached to a system also have the ability to attract or drive people away.

To test the effect of metaphor on pre-use intention to adopt, and pre-use desire to cooperate with an AI system, we ran a third study. In this study, we present participants with AI systems, described using conceptual metaphors. We ask participants to identify which systems they are more likely to try out and potentially adopt, prior to using the system. 

\subsection{Procedure}
We perform a between-subjects experiment. Each participant was introduced to an AI agent described using one of the metaphors in Study 1. 
Unlike the previous experiments, participants do not actually interact with an AI system (or wizard). Instead, they are asked to rate their likelihood of trying out a new AI system service described by each metaphor.

\subsection{Measures}
To probe for the participants' intentions to try the system, we asked them the following two questions on $5$ point Likert scales: \textit{How likely are you to try out this AI system?}, and \textit{Do you envision yourself engaging in long-term use of such a AI system?} These two questions are combined to form a trial index ($\alpha = 0.94$). 

To understand the participants’ pre-use desire to cooperate with the system, they are asked on a $5$ point Likert scale: \textit{How likely are you to cooperate with such an AI system?}, and \textit{How likely are you to tolerate errors made by this AI system?} These two questions are combined to form a pre-use desire to cooperate index ($\alpha = 0.78$).

\subsection{Participants}
Similar to the previous studies, we recruited participants from AMT. $80$ new participants participated in this survey: $20$ participants exposed to a metaphor from each quadrant. We ensured that none of the participants in this study participate in any of our other studies.

\subsection{More interest in trying out high competence and high warmth AI systems}
Participants were more interested to try out AI systems that were described by high competence and high warmth. A two-way ANOVA revealed that competence $(F(1, 100) = 42.96, p < 0.001, \eta^2 = 0.34)$ and warmth $(F(1, 100) = 5.63, p = 0.020, \eta^2 = 0.04)$ both had significant impact on their intention to try out the AI system. The average trial index response was $(2.95 \pm 1.18)$ for high competence, and $(1.5 \pm 0.83)$ for low competence. Following a similar pattern, the average trial index response was $(2.48 \pm 1.28)$ for high warmth as opposed to $(1.96 \pm 1.16)$ for low warmth. The ANOVA also showed an interaction effect between competence and warmth $(F(1, 100) = 1.54, p = 0.021, \eta^2 = 0.012)$. In this case, the combination of high competence and high warmth produced a substantial benefit $3.35 \pm 1.06$ compared to the effects of warmth and competence individually: low competence and high warmth $1.63 \pm 0.84$, high competence and low warmth $2.55 \pm 1.77$, and both low competence and low warmth $1.38 \pm 0.82$.

Participants were more likely to cooperate positively towards AI systems that were described by high competence and high warmth. A two-way ANOVA revealed that competence $(F(1, 100) = 37.36, p < 0.001, \eta^2 = 0.28)$ and warmth $(F(1, 100) = 5.63, p = 0.01, \eta^2 = 0.08)$ both had significant impact on the trial index. The average pre-use desire to cooperate index response was $(2.80 \pm 1.11)$ for the high competence as opposed to $(1.58 \pm 0.84)$ for low competence. Similarly, the average pre-use desire to cooperate index response was $(2.53 \pm 1.18)$ for the high warmth as opposed to $(1.85 \pm 1.04)$ for low warmth. The ANOVA also showed an interaction effect between competence and warmth $(F(1, 100) = 6.86, p = 0.001, \eta^2 = 0.052)$. People expected to behave more positively with high competence and high warmth AI systems $3.40 \pm 0.75$ over the low competence and high warmth $1.65 \pm 0.82$, high competence and low warmth $2.20 \pm 1.09$, and both low competence and low warmth $1.50 \pm 0.87$ bots.

\subsection{Summary}
While our previous studies demonstrated the detrimental effects of presenting an AI system with a high competence metaphor, this study shows a positive benefit of high competence --- people are more likely to try out a new service if it is described with a high competence metaphor. This study also shows that metaphors that project high warmth also increase people's likelihood of trying out a service and to behave positively with it. We discuss the implications of of these findings and suggest guidelines for choosing metaphors considering both the competing objectives of attracting more users and ensuring favorable evaluations and cooperative behavior.

\section{Discussion}
Metaphors, as an expectation setting mechanism, are task- and model-agnostic. Users reason about complex algorithmic systems, including news feeds~\cite{DeVito:2018:PFF:3173574.3173694}, content curation, and recommender systems, using metaphors. This implies their effects are not limited to conversational agents or even to AI systems and can be used to set expectations of any algorithmic system (e.g., is Facebook's newsfeed algorithm a gossipy teen, an information butler, or a spy?), although the implications of our study might differ depending on the task, interaction and context. 

With our findings in mind, this section explores their design implications and limitations, and situates our work amongst existing literature in HCI. We end with a retrospective analysis on existing and previous conversational AI products, reinterpreting their metaphors and adoption/user cooperation patterns through the lens of our results.

\subsection{User behavior around algorithmic systems}
Our work contributes to a growing body of work in HCI that seeks to understand how people reason about algorithmic systems with the aim of facilitating more informed and engaging interactions~\cite{french2017s,DeVito:2018:PFF:3173574.3173694,eslami2015}. Previous work has looked at how users form informal theories about the technical mechanisms behind of social media feeds~\cite{french2017s,DeVito:2018:PFF:3173574.3173694,eslami2016} and how these ``folk theories'' drive their interactions with these systems. People's conceptual understanding of such systems have been known to be metaphorical in nature, leading them to form folk theories of socio-technical systems in terms of metaphors.
Folk theories for Facebook and Twitter news feeds include metaphors rooted in personas such as ``rational assistant'' and ``unwanted observer'' as well as metaphors tied to more abstract concepts such as ``corporate black box'' and ``transparent platform''. More recent work has sought to study the social roles of algorithms by looking at how people personify algorithms and attach personas to them~\cite{wu2019agent}. Prior work in the domain of interactive systems and embodied agents has observed that the mental schemas people apply towards agents affect the way they behave with the agent and it is possible to detect users' schematic orientation through initial interaction~\cite{lee2010receptionist}. Diverging from previous work on folk theories, our work takes a complementary route --- instead of studying \textit{which} metaphors users attach to systems, we study \textit{how} metaphors explicitly attached to the system, by designers, impact experiences.

\subsection{Design implications}
Studies 1 and 2 demonstrate that low competence metaphors lead to the highest evaluations of an AI agent, but Study 3 counters that agents with low competence metaphors are least likely to be tried out. What should a designer do?

From Study 3, it becomes clear that associating a high warmth metaphor is always beneficial---however, the choice of competence level projected by the metaphor becomes a more nuanced decision. 
One possible approach might be to choose a higher-competence metaphor but to lower competence expectations right after interaction begins (e.g., ``Great question! One thing I should mention: I'm still learning how to best respond to questions like yours, so please have patience if I get something wrong.'')
Another approach might be to age the metaphor over time: to present a high competence metaphor such as a professional, but when a user first encounters it, the agent introduces itself via a lower-competence version such as a professional trainee and tells the user that it will evolve over time into a full professional~\cite{seeringtakes}.

If designers are unwilling to change or adapt their high-competence metaphor, then their designs run the risk of being abandoned for being less effective than users expect. There may be other ways to disarm the contrast between expectations and reality. The agent blaming itself for errors or blaming the user for errors create challenging issues, but blaming an intermediary might work~\cite{nass2005wired}: for example, ``I've seen that previous folks who asked that question meant multiple different things by it. To make sure I can help effectively, can you reword that question?''

\subsection{Limitations and future work}
The scope of the study was limited to a conversational AI, as an instance of an algorithmic system, where interaction is devoid of strong visual cues~\cite{saygin2011thing}. In the case of embodied agents and systems where visual communication is a major aspect of the interaction, visual factors might have a strong effect on expectations. It is important to understand how users factor in these visual signals in forming an impression of the system. Additionally, our choice of conceptual metaphors was solely textual metaphors; future work should explore how these findings translate to visual metaphors such as the abstract shape associated with Siri, or the cartoonish rendering of Clippy, because such visual abstractions also inform users' judgements of a system's competence and warmth.

Since the task in our study was highly structured and participants had no incentive to explore peripheral conversational topics, we did not observe significant differences in user vocabulary across the conditions. This result surprised us --- that evaluations would differ even if the interactions themselves had no major differences between conditions. Future work should explore user behavior in open-ended conversations, which are more likely to contain personal stories and anecdotes that can elicit greater behavior changes. The conversations and therefore, interactions with the AI system were limited to $15-20$ minutes, so further research needs to establish the effects of metaphors on prolonged exposure to the AI system. Additionally, our service is not commonly used by people today to book flights or hotels and it is possible that the novelty of performing this task with a conversational agent might have skewed evaluations. This necessitates the need to understand how prior experience with similar technology changes people's susceptibility to such expectation shaping and subsequently their evaluations.

We observed partial support for Assimilation Theory along the warmth axis: participants preferred to cooperate with agents projecting higher warmth but at the same time, they perceived a difference between the agent's projected warmth and the actual warmth. Future work is needed to develop more robust theories along the warmth axis. One potential direction could create conditions of more extreme violation --- sampling metaphors that signal either extremely low or extremely high warmth and measuring to see if participants' attitudes towards the agent are still driven by their pre-use warmth perceptions or whether the larger perceived difference in warmth alters their attitudes towards the agent.

Our study explored the effect of metaphors on adoption and behavioral intentions but these could also impact many other factors, including perceived trustworthiness of the system.
Along this direction, future work should explore the impact on user evaluations when the interaction with the AI system results in a failure to accomplish the task. Finally, the actually competence and warmth of the AI system should be varied to analyze the effects metaphors as the AI system's competence is lowered from our human-level performance.

\subsection{Retrospective analysis}
Studies have repeatedly shown that initial user expectations of their conversational agents (including Google Now, Siri, and Cortana) are not met \cite{Zamora:2017:ISD:3125739.3125766,jain2018evaluating,luger2016like}, causing users to abandon these services after initial use. Initial experiences with conversational agents are often decisive: users reported that initial failures in achieving a task with Siri caused them to retreat to simple tasks they were sure the system could handle. While bloated expectations of users before interacting with AI systems have been acknowledged, little work has explored what those expectations are and how they contribute to user adoption and behavior.

Are today's conversational agents being set up for failure? Our studies establish that the descriptions and metaphors attached to these systems can play a key role in shaping expectations. Woebot was introduced as ``Tiny conversations to feel your best''; Replika presaged as ``The AI companion who cares'' and Mitsuku was revealed as ``a record breaking five-time winner of the Loebner Prize Turing Test [...] the world's best conversational chatbot''. We collected these descriptions associated with $5$ popular social chatbots---Xiaoice, Mistuku, Tay, Replika and Woebot---and deployed the exact same warmth-competence measurement of those descriptions with $50$ participants from AMT as we used for the metaphors in our study. 

 \begin{figure}[t]
    \centering
    \includegraphics[width=0.8\textwidth]{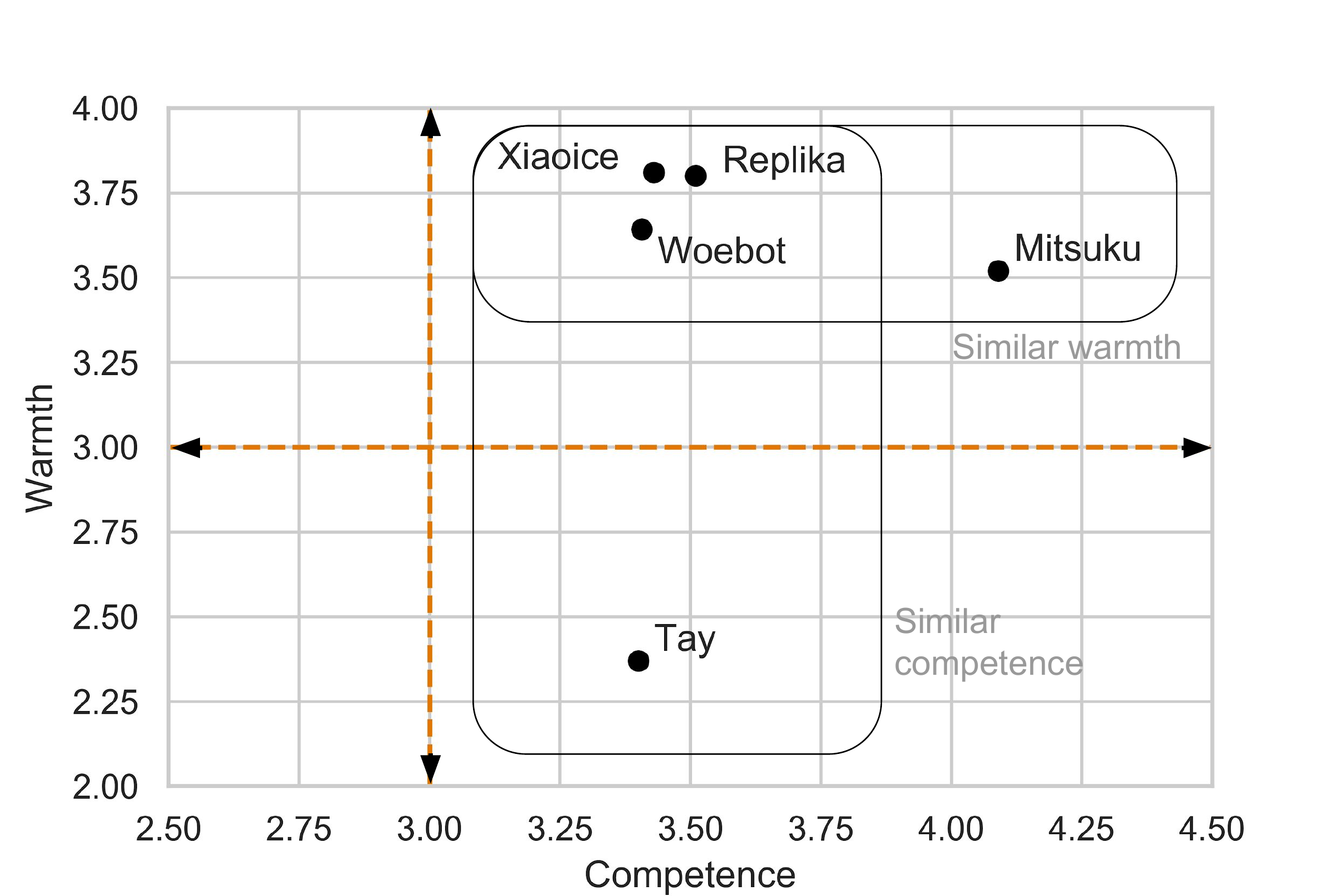}
    \caption{Average warmth and competence measured for popular social chat-bots. Both axes range from $1$ to $5$.}
    \label{fig:bots}
\end{figure}

We find that today's social chatbots signal high competence (Figure~\ref{fig:bots}), between ``recent graduate'' and ``trained professional''. (Tay, incidentally, also projects very low warmth.) Descriptions of this kind, as we've shown, might be setting such systems up for failure. As Ars Technica reported: ``You might come away thinking that Apple found some way to shrink Mad Men's Joan Holloway and pop her into a computer chip. Though Siri shows real potential, these kinds of high expectations are bound to be disappointed''~\cite{Cheng}. It is important to note, then, that users often report disappointment after using these agents, especially since Apple's announcement of Siri included the sentence: ``Ask Siri and get the answer back almost instantly without having to type a single character''; Google Assistant was heralded as ``the first virtual assistant that truly anticipates your needs''.

With the recent glut of Twitter bots and other social agents that learn from their interactions and are adaptive in nature~\cite{park2019ai}, it also becomes important to understand what drives users' antisocial behaviour towards such bots and what factors contribute to antisocial behavior. Previous work has sought explanations through the lens of user profiling, gender attributions, and racial representations. Our work provides another lens on why otherwise similar systems such as Xiaoice and Tay (both female, teen-aged, and not representative of marginalised communities in their respective countries) might have elicited vastly different responses from their users. While Tay's official Twitter account described it as ``Microsoft's AI fam from the internet that's got zero chill!'', signaling high competence and low warmth, Xiaoice was setup to be ``Sympathetic ear'' and an ``Empathetic social chatbot''~\cite{DBLP:journals/corr/abs-1812-08989}, very clearly signaling high warmth and even priming behaviors around warmth such as personal disclosure. Our study suggests that people are more likely to cooperate with a bot that is perceived as higher warmth before use --- a result consistent with the fact that Xiaoice continued to be a friend and remained popular with its user base while Tay was pulled down within $16$ hours of its release for attracting trolls. 

Xiaoice is not an isolated case. Other bots, such as Woebot and Replika, which were set up as with high warmth, have had success in garnering users. Even though both Woebot and Replika had comparable competence expectations to that of Tay's, they were far warmer and obtained an altogether different outcome. Similarly, among the bots perceived as high warmth, Mitsuku stood out as as exceptionally competent and consistent with our finding that perceptions of very high competence decrease the desire to cooperate with the AI system: up to $30\%$ of the messages received by Mitsuku comprise of antisocial messages~\cite{Worswick}. For their part, Microsoft may have absorbed the lesson, as Tay's successor named Zo, was described more warmly as ``Always down to chat. Will make you LOL''.

While we acknowledge that there are several variables that affect user reception of these systems, the fact that our findings are consistent with in-the-wild outcomes of extant conversational systems is notable. It is, of course, impossible to prove that the expectations set by attached metaphors are a causal factor in the users' reception of these specific systems, and caution readers against concluding that metaphors alone are responsible.
\section{Conclusion}
We explore metaphors as a causal factor in determining users' evaluations of AI agents. We demonstrate experimentally that these conceptual metaphors change users' pre-use expectations as well as their post-use evaluations of the system, their intentions of adopt and their desire to cooperate. While people are more likely to cooperate with agents that they expect to be warm, they are more likely to adopt and cooperate with agents that project low competence. This result runs counter to designers' usual default towards projecting high competence to attract more users.

\begin{acks}
We thank Jacob Ritchie, Mitchell Gordon and Mark Whiting for their valuable comments and feedback. This work was partially funded by the Brown Institute of Media Innovation and by Toyota Research Institute (TRI) but this article solely reflects the opinions and conclusions of its authors and not TRI or any other Toyota entity.
\end{acks}

\bibliographystyle{ACM-Reference-Format}
\bibliography{bibliography}

\end{document}